\documentclass[prd,twocolumn,showpacs,showkeys,preprintnumbers,floatfix,
nofootinbib, superscriptaddress]{revtex4}
\usepackage{amsfonts} 
\usepackage{amssymb} 
\usepackage{amsmath} 
\usepackage{graphicx} 
\usepackage{subfigure} 
\usepackage{array} 
\usepackage{dcolumn} 
\usepackage{bm} 
\usepackage{latexsym} 
\usepackage{longtable} 
\usepackage{hyperref} 
\graphicspath{{./figs/}}
\renewcommand{\tilde}{\widetilde}

\newcommand{\CD}{{\cal D}}

\newcommand{\CT}{{\cal T}}

\newcommand{\CM}{{\cal M}}

\begin{document}


\title{One-loop matching factors for
  staggered bilinear operators with improved gauge actions}
\author{Jongjeong Kim}
\email[Email: ]{rvanguard@phya.snu.ac.kr}
\affiliation{
  Lattice Gauge Theory Research Center and 
  Frontier Physics Research Division, \\
  Department of Physics and Astronomy,
  Seoul National University, 
  Seoul, 151-747, South Korea
}
\author{Weonjong Lee}
\email[Email: ]{wlee@snu.ac.kr}
\homepage[Home page: ]{http://lgt.snu.ac.kr/}
%
%
\affiliation{
  Lattice Gauge Theory Research Center and 
  Frontier Physics Research Division, \\
  Department of Physics and Astronomy,
  Seoul National University, 
  Seoul, 151-747, South Korea
}
\author{Stephen R. Sharpe}
\email[Email: ]{sharpe@phys.washington.edu}
%
%
\affiliation{
  Physics Department, Box 351560,
  University of Washington,
  Seattle, WA 98195-1560, USA
}
\date{\today}
\begin{abstract}
We present results for one-loop perturbative matching factors using
bilinear operators composed of improved staggered fermions, using
unimproved (Wilson) and improved (Symanzik, Iwasaki, and DBW2)
gluon actions.  
We consider two fermions
actions---HYP/$\overline{\text{Fat7}}$-smeared and ``asqtad''.
The former is being used in calculations of electroweak matrix
elements, while the latter have been used extensively by the MILC
collaboration.
We observe that using the improved gluon action leads to
small reductions in the perturbative corrections,
but that these reductions are smaller than those obtained
when moving from the tadpole-improved naive staggered action
to either HYP-smeared or asqtad action.
\end{abstract}
\maketitle

\section{Introduction \label{sec:intr}}

Improved staggered fermions are an attractive choice for the numerical
study of QCD, and are being used for a variety of calculations
relevant to phenomenology. For calculations of electroweak matrix
elements, such as our ongoing calculation of 
$B_K$~\cite{wlee-2009-1,wlee-2009-2,wlee-2009-3,wlee-2009-4},
one needs to match continuum
operators in the effective Hamiltonian onto corresponding lattice-regularized 
operators. Here we calculate such matching factors for fermion
bilinears composed of improved staggered fermions with various
gluon actions. We work at one-loop level in perturbation theory.

The motivation for this work is three-fold.
First, the results are a step on the way to the calculation
of matching factors for four-fermion operators, such as that needed
for $B_K$, results for which will be presented in an upcoming
work~\cite{four-fermion-in-preparation}.
Second, our results allow us to compare the efficacy of improvements to
fermion and gauge actions at reducing matching factors.
Third, some of our results can be compared to ongoing 
calculations of matching factors~\cite{AndrewLat09} using 
non-perturbative renormalization (NPR)~\cite{NPR}.
We can also check our result for the mass-renormalization
for asqtad fermions with that obtained (as a byproduct
of a two-loop calculation) in Ref.~\cite{mason-2005-1}.

Two major problems with unimproved staggered fermions are
large taste-symmetry breaking and large perturbative corrections
to matching factors.
Previous work has shown that both problems are
alleviated by smearing the gauge links to which the fermions
couple. In particular, it turns out
that HYP smearing\cite{hasenfratz-01}\footnote{%
At one-loop order, HYP smearing, 
with parameters set to their perturbatively
improved values, is equivalent to using the
$\overline{\text{Fat7}}$ links introduced in Ref.~\cite{wlee-02-1}.
We refer to these simply as HYP links in the
following. The two smearings differ at higher-order and
non-perturbatively.
}
is most effective at reducing one-loop perturbative 
corrections~\cite{wlee-02}, and also in reducing the taste symmetry-breaking
in the pion spectrum~\cite{wlee-08-1,wlee-08-2}.
In light of this we are using such smearing for valence quarks
in our ongoing calculations of matrix elements.
These calculations make use, however, of the MILC 
configurations~\cite{MILC-review-2009},
which use a Symanzik-improved gauge action.
Thus we have undertaken the extension of
the calculation of matching factors to the improved gluon action.
We have done so using both the HYP-smeared action but also
using the asqtad action. The latter gives further information
on the comparison between smearing methods.

The generalization to an improved gluon action
is non-trivial. The gluon propagator is diagonal with the Wilson gauge action
(in the Feynman gauge),
but becomes a full $4\times 4$ matrix for an improved gauge action. 
Thus various simplifications that are possible with the Wilson gauge
action do not occur with the improved gauge action.

A calculation along similar lines has been done previously 
in Ref.~\cite{gamiz-05}. The authors consider the asqtad
fermion action and Symanzik-improved glue, but use
staggered fermion operators containing unsmeared (``thin'') links.
This is in contrast to the operators which we use, in which
all links are smeared.
The paper is organized as follows.
In Sec.~\ref{sec:act-op}, we explain our notation and 
conventions for actions and operators.
In Sec.~\ref{sec:renorm}, we describe the renormalization of
bilinear operators on the lattice.
In Sec.~\ref{sec:matching}, we explain the procedure of matching
between the continuum and lattice operators.
In Sec.~\ref{sec:conclude}, we close with a discussion of
our numerical results.
We relegate technical details to three appendices.
Appendix~\ref{app:sec:imp-glu} discusses the gluon
propagator for improved actions,
App.~\ref{app:one-loop-HYP} gives results
for the renormalization of HYP fermions,
and App.~\ref{app:one-loop-asqtad} describes the
results for asqtad fermions.

A preliminary account of this work has appeared in 
Ref.~\cite{kim-lee-sharpe-lat09}.

\section{Actions, Feynman rules and Operators \label{sec:act-op}}
A general form for the $O(a^2)$-improved gluon action is~\cite{weisz-83}
(using the labeling convention of Ref.~\cite{luscher-84})
\begin{eqnarray}
  S_g &=& \frac{6}{g_0^2}\Bigg[
    c_{\rm 0}\sum_{\rm pl}\frac13 {\rm ReTr} (1 - U_{\rm pl})
  \nonumber\\
  &+&\quad c_{\rm 1}\sum_{\rm rt}\frac13 {\rm ReTr} (1 - U_{\rm rt}) 
  \nonumber\\
  &+&\quad c_{\rm 2}\sum_{\rm pg}\frac13 {\rm ReTr} (1 - U_{\rm pg}) 
  \nonumber\\
  &+&\quad c_{\rm 3}\sum_{\rm ch}\frac13 {\rm ReTr} (1 - U_{\rm ch})\Bigg]\,.
    \label{eq:ImpGluAction}
\end{eqnarray}
%
Here pl, rt, pg, and ch denote the shape of the Wilson 
loops---plaquette, rectangle, parallelogram and chair, respectively.
The overall normalization 
of the coefficients is such that
\begin{equation}
c_0+8c_1+8c_2 + 16 c_3 = 1\,.
\label{eq:c_i-normalization}
\end{equation}
If we consider only on-shell improvement, then
one operator is redundant~\cite{luscher-84},
and we adopt henceforth the convention of setting $c_3=0$.

The Wilson gauge action corresponds to the choices
$c_0=1$, $c_1= c_2=0$.
As shown by L\"uscher and Weisz, tree-level on-shell 
Symanzik-improvement 
of the pure-gluon theory is obtained if
the improvement coefficients takes the values~\cite{weisz-83,luscher-84}
\begin{equation}
c_{0}=\frac53\,,\
c_{1}=-\frac1{12}\,,\
c_{2} = 0\,.
\label{eq:tree-level-Symanzik}
\end{equation}
The MILC collaboration use a one-loop improved action
determined in Refs.~\cite{luscher-85,lepage-95}.
In a perturbative calculation, however, the one-loop corrections
to the improvement coefficients enter at
two-loop order in a calculation of bilinear matching 
coefficients. Thus, for our one-loop calculation of 
matching coefficients, the consistent choice is to
use the tree-level coefficients (\ref{eq:tree-level-Symanzik})
when determining the gluon propagator.

The propagator for the improved gluon action
is well known.
We have found a relatively simple form for this propagator,
which is given in Appendix~\ref{app:sec:imp-glu}.

The staggered fermion actions that we
consider in this paper are the asqtad and HYP
actions. The former is
\begin{widetext}
\begin{eqnarray}
S_\text{asqtad} &=& 
\sum_{n} \bigg[ \bar{\chi}(n)
\sum_{\mu} \eta_{\mu}(n) \Big( 
\nabla_\mu^\text{F7L} \chi(n)
+ \frac{1}{8} [\nabla_\mu^\text{T1} - \nabla_\mu^\text{T3}] 
\chi(n) \Big) 
+ (m/u_0) \bar{\chi}(n)\chi(n) \bigg] \,,
\label{eq:asqtad}
\\
\nabla_\mu^\text{F7L} \chi(n) &=&\frac{1}{2}
[W_\mu (n) \chi(n + \hat{\mu}) - 
W^{\dagger}_{\mu}(n-\hat{\mu}) \chi(n-\hat{\mu})]\,,
\label{eq:fat7pluslepage}
\\
\nabla_\mu^\text{T1} \chi(n) &=&\frac{1}{2 u_0}
[ U_\mu(n) \chi(n + \hat{\mu}) - 
U^\dagger_\mu(n-\hat{\mu})\chi(n-\hat{\mu}) ]\,,
\\
\nabla_\mu^\text{T3} \chi(n) &=&\frac{1}{6 u_0^3}
[U(n,n+3\hat{\mu}) \chi(n + 3\hat{\mu}) - 
U(n,n-3\hat{\mu})\chi(n-3\hat{\mu})]\,,
\end{eqnarray}
\end{widetext}
where $n = (n_1,n_2,n_3,n_4)$ labels lattice sites,
$\eta_{\mu}(n) = (-1)^{n_1 + \cdots + n_{\mu-1}} $ is
the usual staggered phase, 
$U_\mu(n)$ is the original, ``thin'' link,
and $m$ is the quark mass in MILC's convention.
Here and in the following we set the
lattice spacing to unity, except where
clarity dictates otherwise.
$W_\mu (n)$ is a smeared link constructed using the Fat7 blocking
transformation~\cite{orginos-99-1,orginos-99-2} combined with Lepage's
prescription~\cite{lepage-99} and tadpole improvement~\cite{lepage-93}.
$U(n,n\pm3\hat{\mu})$ are products of 3 thin links in the $\mu$
direction,
\begin{eqnarray*}
U(n,n+3\hat{\mu}) &=& U_\mu(n) U_\mu(n+\hat{\mu}) U_\mu(n+2\hat{\mu}) 
\\
U(n,n-3\hat{\mu})&=& U^\dagger_\mu(n-\hat{\mu}) 
U^\dagger_\mu(n-2\hat{\mu}) U^\dagger_\mu(n-3\hat{\mu}) \,,
\end{eqnarray*}
and appear in the Naik term.
Finally, $u_0$ is the tadpole improvement factor, which we
determine here as the fourth-root of the average plaquette.
This action is tree-level $O(a^2)$ improved.

The HYP action is simply the unimproved staggered
action using HYP-smeared links:
\begin{eqnarray}
S_\text{HYP} &=& 
\sum_{n} \bar{\chi}(n) \Big[
\sum_{\mu} \eta_{\mu}(n) \nabla_\mu^\text{H} + m \Big]\chi(n) \,,
\label{eq:SHYP}
\\
\nabla_\mu^\text{H} \chi(n) &=&
\frac12
[V_\mu (n) \chi(n + \hat{\mu}) -
V^{\dagger}_{\mu}(n-\hat{\mu}) \chi(n-\hat{\mu})]
\nonumber
\end{eqnarray}
where $V_\mu$ is constructed using the HYP blocking
transformation of Ref.~\cite{hasenfratz-01}.
This transformation has the advantage of using only
links lying within hypercubes attached to the original
thin-link, so that $V_\mu$ is less extended than the fat links
$W_\mu$ used in the asqtad action.
HYP-blocking also includes SU(3) projection.
We set the HYP blocking parameters to the values
that remove the tree-level coupling of quarks to
gluons having one or more components of momenta equal to $\pi/a$
(with the other components vanishing).
In the notation of Ref.~\cite{wlee-03,wlee-02},
these are the HYP(II) parameters. These are the values
we have used in our simulations.

The HYP action is only partially improved---taste-breaking 
$O(a^2)$ interactions are removed, but taste-conserving $O(a^2)$
terms are not.
It would thus seem to be a poorer choice than
the asqtad action, which is fully $O(a^2)$ improved. 
It turns out, however, to be a better choice
in practice, for two reasons. 
The most important is that it leads to
substantially smaller taste-splittings between pions~\cite{wlee-08-2}.
Since taste-splitting is the dominant $O(a^2)$ effect with
staggered fermions, this means that HYP-smeared quarks have smaller
$O(a^2)$ effects than asqtad quarks.
The second reason is that the HYP action is more continuum-like,
in the sense that loop contributions
to matching factors are typically smaller. This is known explicitly at
one-loop for bilinears (as found for the Wilson gauge action 
in Ref.~\cite{wlee-02},
and for the improved gauge action in the present work---see 
Table~\ref{tab:spread}),
and is expected also to hold at higher order because the four-fermion
operators induced at one-loop have greatly reduced coefficients
compared to asqtad quarks~\cite{HISQ}.
These advantages, as well as the computational simplicity of
implementing this action for valence quarks,
have led us to pursue calculations using the HYP-smeared action.
For completeness,
we note that similar reductions in taste-splittings 
(and presumably similar reductions in one and higher-loop
contributions to matching factors) can also be
obtained using the more highly improved,
and more complicated, HISQ action~\cite{HISQ}.

The thin links are related to the gauge fields $A_\mu$
in the usual way,
\begin{equation}
U_\mu(x) = \exp
\left[ i g_0 A_\mu ( x \!+\! {\hat\mu}/{2}) \right]\,.
\label{eq:thin:gauge}
\end{equation}
Writing the HYP links in a similar way
in terms of  ``blocked gauge fields'' $B_\mu$,
\begin{equation}
V_\mu(x) = \exp
\left[ i g_0 B_\mu ( x \!+\! \frac{\hat\mu}{2}) \right] \,,
\label{eq:fat:gauge}
\end{equation}
the $B_\mu$ can be expressed in terms of the
gauge fields as
\begin{equation}
B_\mu = \sum_{n=1}^{\infty} B_\mu^{(n)}(A_\nu)\,.
\end{equation}
Here $B^{(n)}$ contains all terms with $n$ powers of $A$.
It turns out that we need only $B_\mu^{(1)}$ in the one-loop
calculation.  
While $B_\mu^{(2)}$ enters in one-loop ``tadpole'' diagrams,
these contributions vanish because of the
SU(3) projection~\cite{Patel-Sharpe-bilin,wlee-02,wlee-02-1}.
Thus all we need is the relationship between
$B^{(1)}_\mu$ and $A_\nu$:
\begin{equation}
B^{(1)}_\mu (k) =  \sum_\nu h_{\mu\nu}(k) A_\nu(k) \,,
\end{equation}
where we have gone over to momentum space.
A convenient general form for the kernel
$h_{\mu\nu}(k)$ is (following Ref.~\cite{wlee-02}, but
using the notation of Ref.~\cite{wlee-03}\footnote{%
The reversal of the indices on $\widetilde G_{\nu,\mu}$ is
intended and follows Ref.~\cite{wlee-03}.})
\begin{subequations}
\label{eq:hmunu}
\begin{eqnarray}
h_{\mu\nu}(k) &=& \delta_{\mu\nu} D_\mu(k) +
(1 - \delta_{\mu\nu})  
{\bar s}_\mu {\bar s}_\nu \widetilde G_{\nu,\mu}(k) \,,
\label{eq:hmunu_a}
\\
D_\mu(k) &=&  1 - d_1 \sum_{\nu\ne\mu} {\bar s}_\nu^2
+ d_2 \sum_{\nu < \rho \atop \nu,\rho\ne\mu}{\bar s}_\nu^2 {\bar s}_\rho^2
\nonumber \\
& & - d_3 {\bar s}_\nu^2 {\bar s}_\rho^2 {\bar s}_\sigma^2
- d_4 \sum_{\nu\ne\mu} {\bar s}_\nu^4 \,,
\\
\widetilde G_{\nu,\mu}(k) &=& d_1
- d_2 \frac{({\bar s}_\rho^2+ {\bar s}_\sigma^2)}{2}
+ d_3 \frac{{\bar s}_\rho^2 {\bar s}_\sigma^2}{3}
+ d_4 {\bar s}_\nu^2 \,.
\end{eqnarray}
\end{subequations}
Here ${\bar s}_\mu = \sin(k_\mu/2)$.

The coefficients $d_i$ distinguish different choices
of smeared links.
\begin{enumerate}
\item[(i)] Unimproved (``thin'') links ($U_\mu$):
\begin{equation}
d_1 = 0, \quad
d_2 = 0, \quad
d_3 = 0, \quad
d_4 = 0.
\end{equation}
\item[(ii)] HYP-smeared fat links ($V_\mu$) whose
  coefficients are chosen to remove $O(a^2)$ taste-symmetry breaking
  at tree level:
\begin{equation}
d_1 = 1, \quad
d_2 = 1, \quad
d_3 = 1, \quad
d_4 = 0.
\end{equation}
\item[(iii)] Fat7 links with the Lepage term ($W_\mu$),
  which remove all $O(a^2)$ couplings at tree level, both
  taste-violating and conserving:
\begin{equation}
d_1 = 0, \quad
d_2 = 1, \quad
d_3 = 1, \quad
d_4 = 1.
\end{equation}

\end{enumerate}

With the exception of some tadpole diagrams,
the perturbative calculation requires the propagator
from one smeared-link to another.
This takes the form
\begin{eqnarray}
\lefteqn{\langle B^{(1),b}_\mu(k) B^{(1),c}_\nu(-k) \rangle}
  \nonumber \\
  &=& \sum_{\alpha,\beta} 
  h_{\mu\alpha}(k) \  h_{\nu\beta}(-k) \
  \langle A^b_\alpha(k)  A^c_\beta(-k) \rangle
  \nonumber \\
  &= & \delta^{bc} \sum_{\alpha,\beta} 
  h_{\mu\alpha}(k) \  h_{\nu\beta}(k)
  D^\text{Imp}_{\alpha\beta}(k) 
\nonumber\\
  &\equiv& \delta^{bc} \CT_{\mu\nu}\,,
\label{eq:smsm_prop}
\end{eqnarray}
where $b,c$ are color indices, and
$D^\text{Imp}_{\mu\nu}$ is the 
propagator for the improved gluon action given in 
eq.~(\ref{eq:ImpGluAction}).
In the third line we have used the fact that $h_{\mu\nu}$ is an even
function of the momentum.
We note that this smeared-smeared propagator includes off-diagonal
($\mu\ne\nu$) terms {\em even if the gluon action is unimproved}, because
the kernel $h_{\mu\nu}$ has off-diagonal terms.
Thus, for those diagrams involving the smeared-smeared propagator,
the generalization to an improved gluon action does not
introduce any new types of contribution, and we can carry
over the form of most of the results from Ref.~\cite{wlee-02}.
Appendix~\ref{app:one-loop-HYP}
describes how this works.

For the asqtad action, however, the situation is less simple,
since not all the links are smeared. Some diagrams must
be calculated anew when using an improved gluon propagator.
We discuss this in Appendix~\ref{app:one-loop-asqtad}.

We now turn to the bilinear operators. We construct them from the
standard hypercube convention~\cite{klu-83}, in which the spin and
tastes of the two continuum fermions are spread over a hypercube:
\begin{widetext}
\begin{eqnarray}
[ S \times F ](y)
& = & \frac{1}{16} \sum_{A,B}
[\bar{\chi_b}(y+A) \
(\overline{ \gamma_S \otimes \xi_F } )_{AB} \
\chi_c(y+B)] \  {\cal V}^{bc}(y+A,y+B) \,.
\label{eq:bilindef}
\end{eqnarray}
\end{widetext}
where $y$ denotes the particular $2^4$ hypercube,
and $A,B$ are ``hypercube vectors'' denoting the
positions within the hypercube.
The matrices $(\overline{ \gamma_S \otimes \xi_F } )_{AB}$ are in the
standard notation of Refs.~\cite{Daniel-Kieu-86,Daniel-Sheard-88}.
The spin ($S$) and taste ($F$) of the bilinear can each be scalar, $ S
$, vector, $ V_\mu $, tensor, $ T_{\mu\nu} $, axial vector, $ A_\mu $,
or pseudoscalar, $ P $.
The only new feature of these operators compared to those used with
unimproved staggered fermions lies in the links used to make them
gauge invariant.
The factor $ {\cal V}^{bc}(y+A,y+B) $ is constructed by averaging over
all of the shortest paths between $ y+A $ and $ y+B $, and for each
path forming the product of {\em smeared gauge links}.
We use the same smeared links as in the fermion action,
i.e. $U_\mu$ for the unimproved action,
$V_\mu$ for the HYP action, 
and $W_\mu$ for the asqtad action.
This ensures the conservation of the current $[V\times S]$
for unimproved and HYP-smeared fermions.

For the asqtad action, however, the presence of the 
three-link (Naik) term in the action means that
$[V\times S]$ is not the conserved current.
As a check, we have also calculated the
matching factor for the asqtad conserved vector current.
This is described in Appendix~\ref{app:one-loop-asqtad}.

Finally, we have also implemented mean-field improvement
for the HYP action~\cite{wlee-02}. 
Although the HYP-smeared links fluctuate
much less than thin links, residual fluctuations are present
and can be partly removed by rescaling the links. 
The rescaling factor, $u_0^{\rm SM}$, is chosen to be the fourth-root
of the plaquette constructed from smeared links.
The details of this procedure are explained in
Ref.~\cite{Patel-Sharpe-bilin} and we do not repeat them
here.

\section{Renormalization of Bilinear Operators 
  \label{sec:renorm}}
\begin{figure}[htbp!]
\subfigure[\ X]{\includegraphics[width=0.23\textwidth]{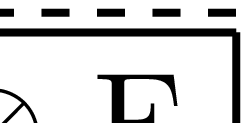}}
\subfigure[\ Y]{\includegraphics[width=0.23\textwidth]{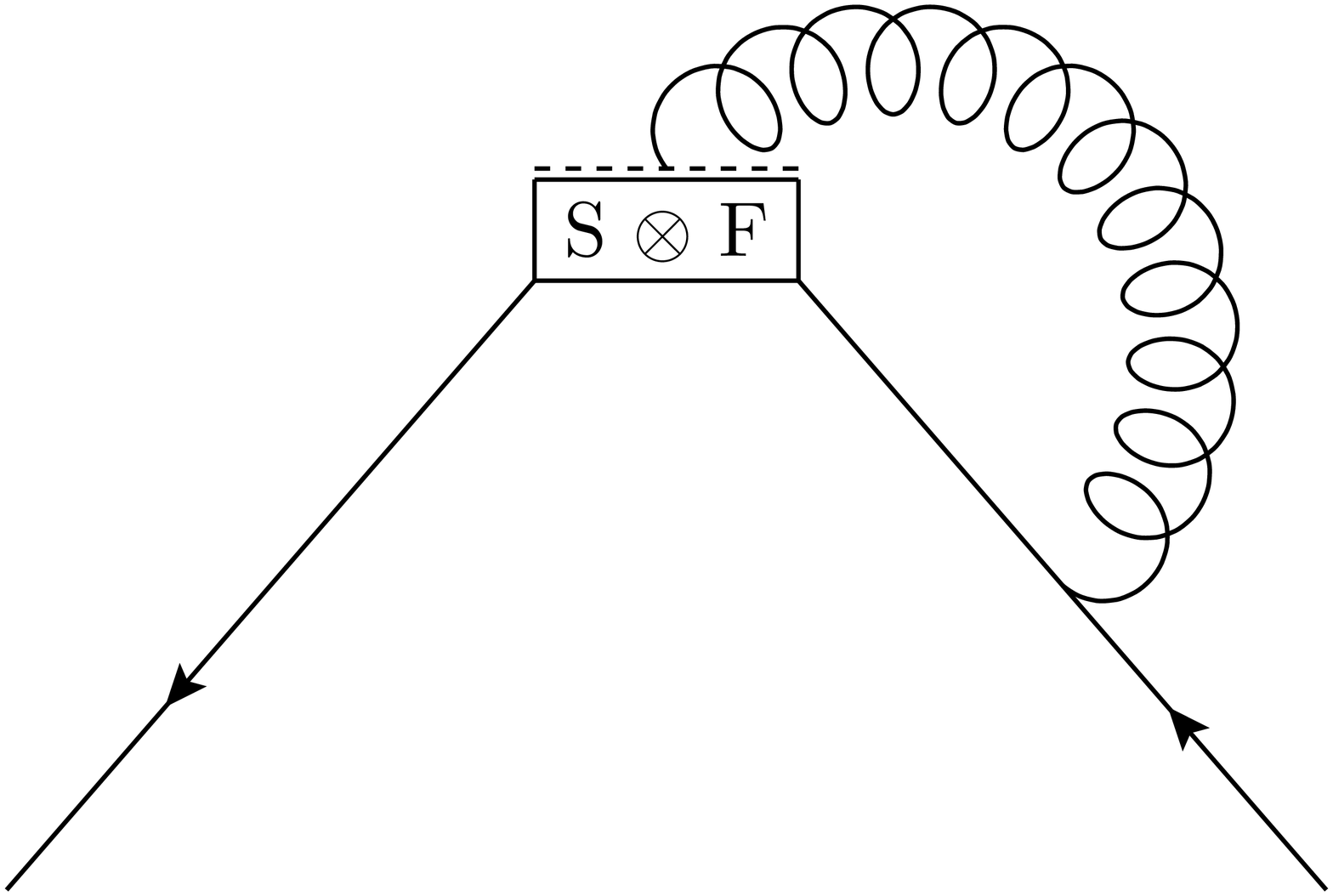}}
\subfigure[\ T]{\includegraphics[width=0.23\textwidth]{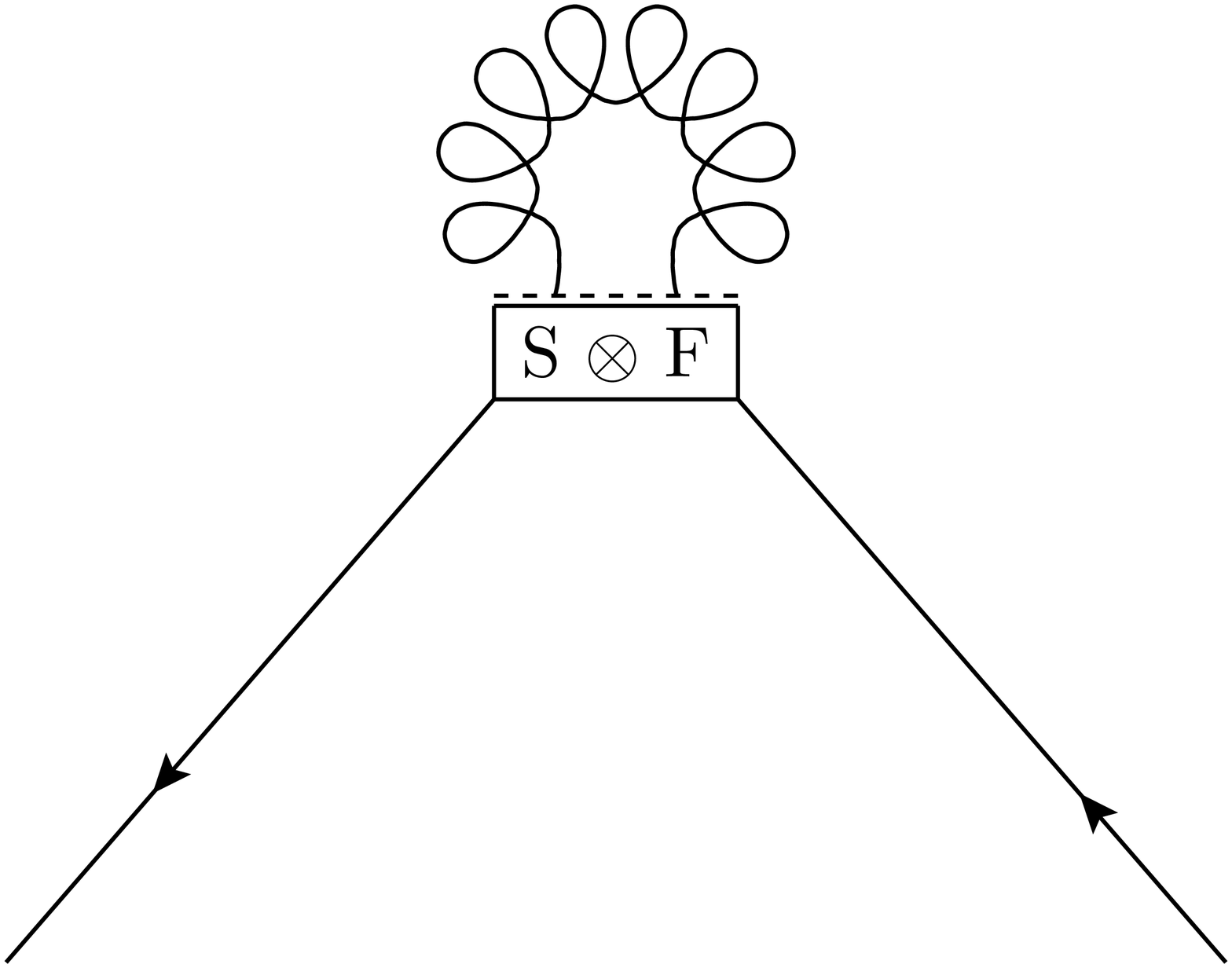}}
\subfigure[\ ZT]{\includegraphics[width=0.23\textwidth]{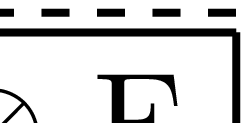}}
\subfigure[\ Z]{\includegraphics[width=0.23\textwidth]{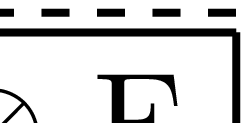}}
\caption{One-loop Feynman diagrams contributing to matching
factors for bilinear operators. 
\label{fig:bi-op}}
\end{figure}

The one-loop Feynman diagrams 
are shown in Fig.~\ref{fig:bi-op}.
The X and Z diagrams are infrared divergent.
We regularize this divergence,
following Refs.~\cite{GS,Patel-Sharpe-bilin} 
by adding a gluon ``mass'' term,
$\lambda^2$, to the denominator of the gluon propagator.
This allows us to set both quark masses and external momenta
to zero.

We have undertaken two independent calculations, one based on the
approach of Ref.~\cite{wlee-95-1},
the other following Refs.~\cite{Patel-Sharpe-bilin,wlee-02}.
These two methods lead to identical results within the
accuracy of the numerical integrations.
We refer to these references for discussions of the methodology.

One-loop matrix elements of lattice operators take the general form
(with the lattice spacing restored for clarity)
\begin{eqnarray}
\lefteqn{{\cal M}^{\text{Latt},(1)}_i =}
\nonumber\\
&& \left\{\delta_{ij} + \frac{g^2}{(4\pi)^2} \Big[
        \delta_{ij} \gamma_{i}^{(0)} \log( a \lambda ) +
        C^{\rm Latt}_{ij}
        \Big]\right\}
        {\cal M}^{{\rm Latt},(0)}_j 
\nonumber \\
& & + O(a)
\label{eq:ff-latt}
\end{eqnarray}
where the superscript indicates the order in perturbation theory,
and the subscript labels the different spins and tastes.
Thus ${\cal M}^{{\rm Latt},(0)}_j$ is tree-level matrix
element of the $j$'th bilinear operator.
$\gamma_{i}^{(0)}$ are the one-loop
anomalous dimensions (which, for bilinears, are diagonal):
\begin{eqnarray}
\gamma_{i}^{(0)} &=&  - 2 C_F d_i\,.
\end{eqnarray}
Here $C_F = 4/3$, while
the $d_i$ depend on the spin, but not on the taste, of the bilinear:
\begin{equation}
d_i = \{3,0,-1,0,3\} \ \ {\rm for}\ \ \{S, V, T, A, P\}\,.
\end{equation}
Finally, $C^\text{\rm Latt}_{ij}$ is the finite part of the correction.

The finite part can be broken down as
\begin{eqnarray}
C^{\rm Latt}_{ij} &=& C_F \delta_{ij} d_i(F_{0000}-\gamma_E+1)
\nonumber\\
&&+C_F\left[X_{ij} + \delta_{ij}\left( 
Y_i + T_i + ZT + Z\right)\right]\,.
\label{eq:Clattform}
\end{eqnarray}
The first line is the finite coefficient accompanying the 
$\log(a\lambda)$, and is thus proportional to the anomalous dimension
matrix. The numerical values of the constants are
$F_{0000}=4.36923(1)$ and $\gamma_E=0.577216\dots$.
The second line gives the finite contributions from each of the
diagrams, and
incorporates the result that only the $X$-diagrams give
rise to mixing between bilinears.

Expressions for the finite contributions are given
in Appendices~\ref{app:one-loop-HYP} (HYP fermions)
and \ref{app:one-loop-asqtad} (asqtad fermions).
We present numerical values in Tables
\ref{tab:Cii-all} (diagonal components)
and
\ref{tab:Cij-all} (off-diagonal components).
Results are shown with both Wilson and improved
gauge actions. 
We quote only two decimal places for brevity; our numerical
evaluations are accurate to at least $1\times 10^{-3}$.
For fermion actions, we compare
the tadpole-improved staggered action 
[i.e. the action of eq.~(\ref{eq:SHYP}) with links $U_\mu/u_0$],
the HYP action, with and without mean-field improvement,
and the asqtad action.\footnote{%
The results with the Wilson gauge action agree with those
obtained in Refs.~\cite{Patel-Sharpe-bilin,wlee-02},
with the exception of the mixing coefficients for
asqtad action [column (g) of Table~\protect\ref{tab:Cij-all}],
where a (numerically small) error in
Ref.~\cite{wlee-02} has been found. We note that
the off-diagonal mixing coefficients in Table~\ref{tab:Cij-all}
must be multiplied by $-3/4$ to be compared to those quoted in
Refs.~\cite{Patel-Sharpe-bilin,wlee-02}.}

We note that improving the gauge action
leads to a moderate decrease in the magnitude of the 
matching coefficients,
except for those which were already small
($|C_{ii}| \lesssim 1$).

\begin{table*}[ht]
\caption{ Diagonal part of finite coefficients,
  $C_{ii}^\text{Latt}$.  Indices $\mu$, $\nu$, $\rho$ and $\sigma$
are all different.  
Results are given for the following choices of fermion and gauge actions: 
$(a)$ Tadpole-improved staggered fermions with Wilson gluon action;
$(b)$ Tadpole-improved staggered fermions with improved gluon action;
$(c)$ HYP fermions with Wilson gluon action; 
$(d)$ HYP fermions with improved gluon action; 
$(e)$ Mean-field improved HYP fermions with Wilson gluon action;
$(f)$ Mean-field improved HYP fermions with improved gluon action;
$(g)$ Asqtad fermions with Wilson gluon action;
$(h)$ Asqtad fermions with improved gluon action.
For brevity, we quote only two decimal places of the numerical results.
\label{tab:Cii-all}
}
\begin{ruledtabular}
\begin{tabular}{lrrrrrrrr}
Operator  & $(a)$ \quad &$(b)$ \quad &$(c)$ \quad &$(d)$ \quad 
& $(e)$ \quad &$(f)$ \quad &$(g)$ \quad &$(h)$ \quad 
\\ 
\hline 
$(1 \otimes 1)$	                               
& 42.47 & 34.12 &  3.46&  2.54&  2.06&  1.58&  6.23 &  4.83  \\
	        $(1\otimes \xi_\mu)$      	    	                  
& 14.86 & 12.28 &  0.05& -0.24&  0.05& -0.24&  3.77 &  2.84  \\
	        $(1\otimes \xi_{\mu\nu})$ 	    	                   
&  2.58 &  2.10 & -3.15& -2.85& -1.75& -1.89&  4.40 &  3.25  \\
	        $(1\otimes \xi_{\mu5})$   	    	
& -3.65 & -3.29 & -6.36& -5.44& -3.55& -3.51&  6.19 &  4.62  \\
	        $(1\otimes \xi_5)$        	    		
& -8.33 & -7.40 & -9.56& -8.01& -5.34& -5.12&  8.37 &  6.34  \\
	        $(\gamma_\mu \otimes 1)$            		            
&  0.00 &  0.00 &  0.00&  0.00&  0.00&  0.00& -1.89 & -1.91  \\
	        $(\gamma_\mu \otimes \xi_\mu)$      		          
&  6.55 &  5.32 &  1.57&  1.21&  0.17&  0.24& -5.70 & -5.17  \\
	        $(\gamma_\mu \otimes \xi_\nu)$      		          
& -0.23 & -0.40 & -2.46& -1.90& -1.06& -0.93&  2.01 &  1.32  \\
	        $(\gamma_\mu \otimes \xi_{\mu\nu})$ 		              
&  4.53 &  3.46 & -0.49& -0.42& -0.48& -0.42& -1.91 & -2.00  \\
	        $(\gamma_\mu \otimes \xi_{\nu\rho})$		
& -3.34 & -3.06 & -4.99& -3.84& -2.18& -1.91&  5.37 &  4.12  \\
	        $(\gamma_\mu \otimes \xi_{\nu5})$   		             
& -0.25 & -0.51 & -2.80& -2.22& -1.40& -1.25&  1.37 &  0.74  \\
	        $(\gamma_\mu \otimes \xi_{\mu5})$   		             
& -6.52 & -5.80 & -7.58& -5.82& -3.36& -2.93&  8.66 &  6.87  \\
	        $(\gamma_\mu \otimes \xi_5)$        		        
& -3.69 & -3.44 & -5.29& -4.13& -2.48& -2.20&  4.77 &  3.58  \\
	        $(\gamma_{\mu\nu}\otimes 1)$        		        
& -1.46 & -1.54 & -2.45& -1.79& -1.05& -0.83&  0.72 &  0.23  \\
	        $(\gamma_{\mu\nu}\otimes\xi_\mu)$   		             
& -0.42 & -0.63 & -0.50& -0.34& -0.50& -0.34& -3.79 & -3.59  \\
	        $(\gamma_{\mu\nu}\otimes\xi_\rho)$  		              
& -3.35 & -3.11 & -4.63& -3.40& -1.82& -1.47&  4.90 &  3.77  \\
	        $(\gamma_{\mu\nu}\otimes\xi_{\mu\nu})$         
& -5.43 & -4.28 &  0.94&  0.76& -0.46& -0.20& -9.68 & -8.50  \\
	        $(\gamma_{\mu\nu}\otimes\xi_{\mu\rho})$        
& -1.04 & -1.18 & -2.46& -1.79& -1.06& -0.83&  0.82 &  0.32  \\
	        $(\gamma_{\mu\nu}\otimes\xi_{\rho\sigma})$     
& -5.92 & -5.26 & -6.92& -5.10& -2.70& -2.21&  8.76 &  7.05  \\
\end{tabular}
\end{ruledtabular}
\end{table*}

\begin{table*}[ht]
\caption{
  Non-vanishing mixing coefficients, $C_{ij}^{\rm Latt}$.
  The notation is as in Table \protect\ref{tab:Cii-all}.
  Note that mean-field improvement does not change off-diagonal coefficients.
  \label{tab:Cij-all}
}
\begin{ruledtabular}
\begin{tabular}{ccrrrrrrrr}
Operator-$i$ & Operator-$j$ & 
$(a)$ \quad &$(b)$ \quad &$(c)/(e)$ \quad &$(d)/(f)$ \quad 
     & $(g)$ \quad &$(h)$ \quad 
\\
\hline 
$(\gamma_\mu\otimes\xi_\nu)$ & $(\gamma_\mu\otimes\xi_\mu)$  
&-4.05  &-3.33 &-0.47 &-0.43 &-1.73 &-1.49  \\
$(\gamma_\mu\otimes\xi_{\mu5})$ & $(\gamma_\mu\otimes\xi_{\nu5})$ 
& 0.86  & 0.81 & 0.34 & 0.33 & 0.73 & 0.68  \\
$(\gamma_\mu\otimes\xi_{\mu\nu5})$ & $(\gamma_\mu\otimes\xi_{\rho\nu5})$ 
& 1.98  & 1.72 & 0.37 & 0.36 & 1.09 & 0.98  \\
$(\gamma_{\mu\nu}\otimes\xi_{\mu5})$ & $(\gamma_{\mu\nu}\otimes\xi_{\rho5})$
& 0.90  & 0.73 &-0.01 &-0.004 & 0.23 & 0.19  \\
\end{tabular}
\end{ruledtabular}
\end{table*}

\section{Matching with continuum operators}
\label{sec:matching}
The continuum operators to which we wish to match are 
\begin{eqnarray}
{\cal O}^\text{Cont}_{\gamma_S \otimes \xi_F}
&=& \bar{Q} (\gamma_S \otimes \xi_F) Q
\nonumber \\
&=& \bar{Q}_{\alpha,a} [\gamma_S]^{\alpha\beta} 
[\xi_F]^{ab} Q_{\beta,b}\,,
\end{eqnarray}
where $Q$ is a four-taste quark field with exact SU(4) flavor
symmetry.

The general form of the one-loop matrix elements
of these continuum operators can be expressed as
\begin{eqnarray}
\lefteqn{{\cal M}^{\text{Cont},(1)}_i =}
\nonumber\\
&&\left\{1 + \frac{g^2}{(4\pi)^2} \Big[
        \gamma^{(0)}_{i} \log( \frac{\lambda}{\mu} ) +
        C^\text{Cont}_{i}
        \Big]\right\}
        {\cal M}^{{\rm Cont},(0)}_i \,.
\label{eq:ff-cont}
\nonumber \\
\end{eqnarray}
Here $C^\text{Cont}_{i}$ is the finite part of the
renormalization factor, which, in the $\overline{\rm MS}$ scheme
using naive dimensional regularization for the gamma matrices, is
\begin{equation}
  C^\text{Cont}_{i} = 
\left\{\frac{10}{3},0,\frac{2}{3}, 0, \frac{10}{3}\right\}
\ {\rm for}\ \{S,V,T,A,P\}\,.
\end{equation}

Now we are ready to match the lattice and continuum operators.
The tree-level matrix elements ${\CM}^{{\rm Latt},(0)}_i$
and ${\CM}^{{\rm Cont},(0)}_i$ are matched by construction.\footnote{%
This requires that one does the unitary change of basis 
to convert from $(\gamma_S\otimes \xi_F)$ to 
$\overline{\overline{(\gamma_S\otimes \xi_F)}}$ matrices,
as explained in Ref.~\cite{Patel-Sharpe-bilin}.}
Equating the one-loop matrix elements in eqs.~(\ref{eq:ff-latt})
and (\ref{eq:ff-cont}) leads to:
\begin{eqnarray}
& & {\cal O}^{\text{Cont},(1)}_i = \sum_{j} 
Z_{ij} {\cal O}^{\text{Latt},(1)}_j
\label{eq:ff-match}
\\ & & 
Z_{ij} = \delta_{ij} + \frac{g^2}{(4\pi)^2} 
  \Big[ - \delta_{ij} \gamma^{(0)}_{i} \log( \mu a ) 
  + \bar{c}_{ij} \Big]  
\end{eqnarray}
where $Z_{ij}$ is the matching factor at the one loop level
and $\bar{c}_{ij}$ is 
\begin{equation}
\bar{c}_{ij} = \Bigl( \delta_{ij} C^\text{Cont}_{i} 
                        - C^\text{Latt}_{ij} \Bigr) \,.
\end{equation}

A partial check of our results with 
the improved gauge action can be made by
comparing to the one-loop result for $Z_m$ given in 
Ref.~\cite{mason-2005-1} (as part of a two-loop calculation):
\begin{equation}
Z_m = 1 + \frac{g^2}{4\pi}\left[0.1188(1) - \frac2\pi \log(\mu a)\right]
\,.
\end{equation}
For staggered fermions, one has an exact relation
\begin{equation}
Z_{S}\equiv Z_{1 \otimes 1} = 1/Z_m\,.
\end{equation}
Thus Ref.~\cite{mason-2005-1} would predict
\begin{equation}
Z_S = 1 + \frac{g^2}{(4\pi)^2} \left[
8 \log(\mu a) + C^{\rm Cont}_S - 4.8262 \right]
\,,
\end{equation}
and thus that $C^{\rm Latt}_{1\otimes 1}=4.8262$.
This agrees with our result, which is given in the first row of column (h)
in Table~\ref{tab:Cii-all}. 

\section{Discussion \label{sec:conclude}}

\begin{table*}[ht]
\caption{
Spread of values for the diagonal finite corrections,
$\bar{c}_{ii}$, both for a given spin (leading to a
scale and scheme dependent result), and between
all operators (setting $\mu a=1$).
Notation for columns is as in Table~\protect\ref{tab:Cii-all}.
  \label{tab:spread}
}
\begin{ruledtabular}
\begin{tabular}{crrrrrrrr}
Spin &
$(a)$ \quad &$(b)$ \quad &$(c)$ \quad &$(d)$ \quad &
$(e)$ \quad &$(f)$ \quad &$(g)$ \quad &$(h)$ \quad 
\\
\hline 
S/P & 50.8 & 41.6  & 13.0 & 10.6 & 7.4 & 6.7 & 4.6 & 3.5 \\
V/A & 13.1  &11.1 &  9.2 & 7.0  & 3.5 & 3.2 &14.4 &12.0 \\
T   &  5.5 & 4.6 &  7.9 & 5.9  & 2.2 & 2.0 &18.4 &15.6 \\
All & 50.8 & 41.6 & 14.5 &  12.6 & 8.9 &  8.7 &19.0 &16.0 \\
\end{tabular}
\end{ruledtabular}
\end{table*}

We can use our results to compare the reduction in the size of one-loop
matching factors achieved by different improvement schemes.
The starting point is tadpole-improved staggered fermions
[column (a) in the Tables].
This comparison is most straightforward for vector (and axial) currents,
since for these the anomalous dimensions vanish, 
so that there is no dependence on renormalization scale or scheme,
and $C_{i}^{\rm Cont}$ vanishes.
Thus, for these operators,
$C_{ij}^{\rm Latt}$ gives a direct measure of the
size of the corrections. 
We note that, for the lattices on
which we are presently simulating (with $a\approx 0.045-0.12\,$fm),
the range of values of $g^2/(16\pi^2)$ is
$0.017-0.026$ (evaluating the coupling at scale $1/a$ in the
$\overline{\rm MS}$ scheme).\footnote{%
A similar range is obtained using $\alpha$ in the V-scheme 
evaluated at the scale $2/a$, which is the close to the
calculated $q^*$ values for $Z_m$~\cite{mason-2005-1}.}
Thus a finite coefficient of
size $|C_{ij}^{\rm Latt}|\approx 5$ corresponds to a 10\% one-loop
correction.

There are eight different tastes of  vector currents in
Table~\ref{tab:Cii-all}, and three in Table~\ref{tab:Cij-all}.
We see that corrections for all actions are of moderate size, with the
largest magnitude being $\approx 9$, so there is not much room for
improvement over the simple tadpole-improved action.
We do note, however, that, for all fermion types,
improving the gauge action does lead to a moderate
reduction in the size of the correction,
except when the magnitude of the coefficient is already of order unity.
For most coefficients the reduction is in the range 10-25\%.
This reduction is, however, smaller than that achievable using mean-field
improved HYP fermions, where the reduction is close to 50\%.
Without mean-field improvement, HYP and asqtad vector
currents turn out to have slightly larger one-loop corrections
than those for tadpole-improved staggered fermions.

Turning now to the operators with anomalous dimensions, we can
remove scale and scheme dependence by considering the
differences between the $\bar{c}_{ii}$ for fixed spin and
varying taste. We list in Table~\ref{tab:spread} the spread of values for
each of the three choices of spins. We also include the
spread of values across all spins and tastes, 
choosing $\mu=1/a$. 
This is a useful measure of the range of corrections,
since the variation with $\mu$ is relatively weak.

We see from the table that, for HYP and asqtad fermions,
improving the gluon action reduces all the spreads, although
by a small amount. Once again, the greatest reduction is
achieved by the mean-field improved HYP operators.
We also see that for HYP operators without mean field improvement,
the largest spread is somewhat smaller than that for asqtad fermions.
This is the most important indicator when considering four-fermion
operators, since, after Fierz transformation, bilinears of
all spins appear.

Finally, it is of interest to see how other choices of
improved gauge action compare to the Symanzik action.
As representative examples we show, in Table~\ref{tab:Iwasaki_DBW2},
results for diagonal coefficients for HYP-fermions
with both the Iwasaki ($c=-0.331$, $c'=0$) and
DBW2 ($c=-1.4067$, $c'=0$)~\cite{DBW2} actions.
We also repeat the results with Wilson and Symanzik gauge
actions for comparison.
We see that, although the changes are relatively small,
both Iwasaki and DBW2 actions lead to smaller coefficients.

\begin{table}[ht]
\caption{ Diagonal part of finite coefficients,
  $C_{ii}^\text{Latt}$, for HYP fermions with
(c) Wilson, (d) Symanzik,
(i) DBW2 and (j) Iwasaki gluon actions.
Other notation as in Table~\protect\ref{tab:Cii-all}.
\label{tab:Iwasaki_DBW2}
}
\begin{ruledtabular}
\begin{tabular}{lrrrr}
Operator  & $(c)$\quad & $(d)$\quad & $(i)$ \quad &$(j)$ \quad 
\\ 
\hline 
$(1 \otimes 1)$	                            &  3.46 &  2.54 & -1.89 &  1.03 \\
$(1\otimes \xi_\mu)$      	    	          &  0.05 & -0.24 & -3.07 & -0.99 \\
$(1\otimes \xi_{\mu\nu})$ 	    	          & -3.15 & -2.85 & -4.18 & -2.90 \\
$(1\otimes \xi_{\mu5})$   	    	          & -6.36 & -5.44 & -5.27 & -4.78 \\
$(1\otimes \xi_5)$        	    		        & -9.56 & -8.01 & -6.34 & -6.62 \\
$(\gamma_\mu \otimes 1)$            		    &  0.00 &  0.00 &  0.00 &  0.00 \\
$(\gamma_\mu \otimes \xi_\mu)$      		    &  1.57 &  1.21 &  0.44 &  0.81 \\
$(\gamma_\mu \otimes \xi_\nu)$      		    & -2.46 & -1.90 & -0.69 & -1.28 \\
$(\gamma_\mu \otimes \xi_{\mu\nu})$ 		    & -0.49 & -0.42 & -0.20 & -0.32 \\
$(\gamma_\mu \otimes \xi_{\nu\rho})$		    & -4.99 & -3.84 & -1.39 & -2.59 \\
$(\gamma_\mu \otimes \xi_{\nu5})$   		    & -2.80 & -2.22 & -0.87 & -1.55 \\
$(\gamma_\mu \otimes \xi_{\mu5})$   		    & -7.58 & -5.82 & -2.10 & -3.92 \\
$(\gamma_\mu \otimes \xi_5)$        		    & -5.29 & -4.13 & -1.56 & -2.84 \\
$(\gamma_{\mu\nu}\otimes 1)$        		    & -2.45 & -1.79 &  0.36 & -0.92 \\
$(\gamma_{\mu\nu}\otimes\xi_\mu)$   		    & -0.50 & -0.34 &  0.83 &  0.01 \\
$(\gamma_{\mu\nu}\otimes\xi_\rho)$  		    & -4.63 & -3.40 & -0.15 & -1.94 \\
$(\gamma_{\mu\nu}\otimes\xi_{\mu\nu})$      &  0.94 &  0.76 &  1.22 &  0.74 \\
$(\gamma_{\mu\nu}\otimes\xi_{\mu\rho})$     & -2.46 & -1.79 &  0.36 & -0.92 \\
$(\gamma_{\mu\nu}\otimes\xi_{\rho\sigma})$  & -6.92 & -5.10 & -0.68 & -3.02 \\
\end{tabular}
\end{ruledtabular}
\end{table}

\begin{acknowledgments}
The research of W.~Lee is supported by the Creative Research
Initiatives program (3348-20090015) of the NRF grant funded by the
Korean government (MEST).
The work of S.~Sharpe is supported in part by the US Department of
Energy under grant
DE-FG02-96ER40956. 
\end{acknowledgments}

\appendix
\section{Improved gluon propagator
  \label{app:sec:imp-glu}}
The improved gluon propagator $D^\text{Imp}_{\mu\nu}$ was 
determined in Ref.~\cite{weisz-83}, and presented in a useful
general form in Ref.~\cite{Aoki-Kayaba-Kuramashi}.
Here we present a simpler form.

The improved gluon action takes the following quadratic form in the
gluon fields after covariant gauge fixing:
  \begin{eqnarray}
    {\cal Q}_{\mu\nu} &\equiv& 
\left([\CD^\text{Imp}]^{-1}\right)_{\mu\nu}
    \\
    &=& 
    \frac{1}{\alpha} \hat{k}^2 {\cal P}_{\mu\nu}
    + f \hat{k}^2 \delta^T_{\mu\nu}
    - c {\cal M}_{\mu\nu} \,,
  \end{eqnarray}
where $\alpha$ is the gauge-fixing parameter,
\begin{equation}
\hat{k}^n \equiv \sum_\mu \left(\hat{k}_\mu\right)^n
\ \ {\rm with}\ \ 
\hat{k}_\mu\equiv 2 {\bar s}_\mu = 2 \sin(k_\mu/2)\,,
\end{equation}
${\cal P}$ is the longitudinal projector
\begin{equation}
      {\cal P}_{\mu\nu} = 
      \frac{\hat{k}_\mu \hat{k}_\nu}{\hat{k}^2}
\quad
\left[{\cal P}^2 = {\cal P}\right]\,,
\end{equation}
$\delta^T$ is the transverse delta-function
\begin{equation}
      \delta^T_{\mu\nu} = 
      \delta_{\mu\nu} - {\cal P}_{\mu\nu}
\quad
\left[{\cal P}\delta^T =0\,,\ (\delta^T)^2=\delta^T\right]\,,
\end{equation}
${\cal M}$ is an auxiliary transverse matrix
\begin{equation}
        {\cal M}_{\mu\nu} = 
        \delta_{\mu\nu} \hat{k}_\mu^2 \hat{k}^2 
        - \hat{k}_\mu^3 \hat{k}_\nu 
        - \hat{k}_\mu \hat{k}_\nu^3
        + \frac{ \hat{k}_\mu \hat{k}_\nu \hat{k}^4 }
        { \hat{k}^2 }\,,
\end{equation}
satisfying ${\cal P}{\cal M}=0$ and $\delta^T{\cal M}=0$,
and the function $f$ is
\begin{equation}
    f = (\omega - c' \hat{k}^2 - c \hat{k}^4 / \hat{k}^2) \,.
\end{equation}
The coefficients $\omega$, $c$ and $c'$ are determined from the
parameters of the improved action [see eq.~(\ref{eq:ImpGluAction})]:
\begin{subequations}
  \label{eq:para-2}
  \begin{eqnarray}
    \omega &=& c_0 + 8 c_1 + 8 c_2 + 16 c_3 
    \\
    c &=& c_1 - c_2 - c_3
    \\
    c' &=& c_2 + c_3  
  \end{eqnarray}
\end{subequations}
In the standard normalization convention $\omega=1$,
and this is the value we use in our perturbative calculation.
We keep $\omega$ as a free parameter, however, since this
allows phenomenological estimates of the impact of using
different variants of the improved action.

The inversion of ${\cal Q}$ is facilitated by observing
that ${\cal M}^3$ is dependent on $\delta^T$, ${\cal M}$
and ${\cal M}^2$, as follows from the Cayley-Hamilton theorem
applied to the three-dimensional transverse space.
The result for the improved propagator is:
\begin{widetext}
\begin{equation}
  \CD^\text{Imp}_{\mu\nu} =
  \alpha \frac{{\cal P}_{\mu\nu}}{\hat{k}^2}
  + \frac{
    \left[
      \hat{k}^2 (\hat{k}^2 - \tilde{c} x_1) +  \tilde{c}^2 x_2 
    \right]
    \delta^T_{\mu\nu}
    + \tilde{c} (\hat{k}^2 - \tilde{c} x_1) {\cal M}_{\mu\nu}
    + \tilde{c}^2 ({\cal M}^2)_{\mu\nu}
  }
  {
  f 
  \left\{ 
    \hat{k}^2
    \left[
      \hat{k}^2 (\hat{k}^2 - \tilde{c} x_1) +  \tilde{c}^2 x_2 
    \right]
    - \tilde{c}^3 x_3
  \right\}
  } \,,
\end{equation}
\end{widetext}
where
\begin{subequations}
\begin{eqnarray}
  \tilde{c} &=& c / f \,, \\
  x_1 &=& 
  {\rm Tr}({\cal M}) 
  = (\hat{k}^2)^2 - \hat{k}^4 
  =  2 \sum_{\mu < \nu} \hat{k}_\mu^2 \hat{k}_\nu^2 \,,  
  \\
  x_2 &=& 
  \left[{\rm Tr}^2({\cal M}) - {\rm Tr}({\cal M}^2)\right]/2 
  \nonumber \\
  &=& \hat{k}^2
    \left[
      \hat{k}^6 - (3/2)\hat{k}^2\hat{k}^4 + (1/2)(\hat{k}^2)^3
    \right]
  \nonumber \\
  &=&  3 \hat{k}^2 
     \sum_{\mu < \nu < \rho} 
     \hat{k}_\mu^2 \hat{k}_\nu^2 \hat{k}_\rho^2 \,, 
  \\
  x_3 &=& 
  \left[
    {\rm Tr}^3({\cal M}) 
    - 3 {\rm Tr}({\cal M}) {\rm Tr}({\cal M}^2)
    + 2 {\rm Tr}({\cal M}^3)
  \right]/6 
  \nonumber \\ 
  &=& \frac{(\hat{k}^2)^2}{6}
    \left[
      (\hat{k}^2)^4 
      + 3 (\hat{k}^4)^2
      - 6 \hat{k}^4 (\hat{k}^2)^2
      + 8 \hat{k}^6 \hat{k}^2
      - 6 \hat{k}^8 
    \right]
  \nonumber \\ 
  &=&  4 (\hat{k}^2)^2 
  \hat{k}_1^2 \hat{k}_2^2 \hat{k}_3^2 \hat{k}_4^2 \,.
\end{eqnarray}
\end{subequations}
We note that ${\cal D}^{\rm Imp}$ is symmetric, and that
its off-diagonal elements are proportional to 
$\hat{k}_\mu\hat{k}_\nu$ multiplied by a function
that is even in each of the components of $\hat{k}$.
Thus it is convenient to write the
propagator as
\begin{equation}
  {\cal D}^\text{Imp}_{\mu\nu} =
\delta_{\mu\nu}\CD^{\rm Imp}_{\mu\mu}
+ (1-\delta_{\mu\nu}) \hat{k}_\mu\hat{k}_\nu
\widetilde \CD^{\rm Imp}_{\mu\nu}\,.
\label{eq:impprop_decomp}
\end{equation}

\section{One-loop results for HYP-smeared fermions
 \label{app:one-loop-HYP}}

In this appendix we present the one-loop expressions
for matching factors for HYP-smeared fermions.
The results are presented in a general way such that
they include also unimproved staggered fermions, as
well as the impact of mean-field improvement.

Key building blocks for these results 
are the diagonal and off-diagonal parts of the 
``smeared-smeared propagator'' (\ref{eq:smsm_prop}), 
which are defined through
\begin{eqnarray}
\CT_{\mu\nu} &=& 
\sum_{\alpha,\beta} h_{\mu\alpha} h_{\nu\beta}
\CD^{\rm Imp}_{\alpha\beta} 
\\
&=&
B \left[\delta_{\mu\nu} P_\mu +
(1-\delta_{\mu\nu})4{\bar s}_\mu{\bar s}_\nu O_{\mu\nu}\right]
\,.
\label{eq:POdef}
\end{eqnarray}
Here $B$ is the boson propagator
\begin{equation}
B = \left[4 \sum_\mu {\bar s}_\mu^2\right]^{-1}
= \left[\hat{k}^2\right]^{-1}\,.
\end{equation}
Explicit expressions for $P_\mu$ and $O_{\mu\nu}$ can be 
obtained using the decompositions
(\ref{eq:impprop_decomp}) and (\ref{eq:hmunu_a})
of the improved gluon propagator and smearing kernel, respectively.
They are, however, uninformative and we do not reproduce them here.

As noted in the main text,
the expressions for one-loop matching factors for
HYP-smeared fermions that are given in Ref.~\cite{wlee-02} 
still hold as long as the $P_\mu$ and $O_{\mu\nu}$
defined above are used. This follows because the
generalized $P_\mu$ and $O_{\mu\nu}$ still satisfy
the property of being symmetric separately in each component
of ${\bar s}_\mu$. This property is used to simplify the
expressions.

We think it useful to repeat the one-loop expressions here,
both for the sake of clarity (since Ref.~\cite{wlee-02}
considered other cases not relevant here), and 
in order to facilitate the
subsequent discussion of the results with asqtad fermions.

The diagonal part of the $X$-diagram contribution is
\begin{eqnarray}
X_{ii} &=& \sum_{\mu,\nu}\int_k
\left[\bar c_\mu^2 P_\mu (s_\nu^N)^2 B F^2 V_i(k) - \frac{B^2}4\right]
(-)^{\bar S_\mu+\bar S_\nu} 
\nonumber\\
&& +2 \sum_{\mu<\nu} \int_k s_\mu s_\mu^N s_\nu s_\nu^N O_{\mu\nu} B F^2 
V_i(k)
\nonumber\\
&& \times \left[1 - (-)^{\bar S_\mu+\bar S_\nu}\right]\,.
\label{eq:XdiagHYP}
\end{eqnarray}
Here the integral is
\begin{equation}
\int_k\equiv 16\pi^2 \prod_\mu \int_{-\pi}^{\pi} 
\frac{dk_\mu}{2\pi}\,.
\end{equation}
The new abbreviations are
$\bar c_\mu=\cos(k_\mu/2)$ and $s_\mu=\sin(k_\mu)$.
For HYP fermions $s_\mu^N=s_\mu$, although this will
not hold for asqtad fermions.
The denominator of the fermion propagator is
\begin{equation}
F = \left[\sum_\mu (s_\mu^N)^2\right]^{-1} \,,
\end{equation}
while the vertex factors are
\begin{equation}
V_i(k) = \prod_\mu \cos\left[k_\mu(S\!-\!F)_\mu\right]
\,,
\end{equation}
with $S_\mu$ and $F_\mu$ being hypercube four-vectors
describing, respectively, the spin and taste of the bilinear
(which are collectively labeled ``$i$'').

The non-zero mixing coefficients are ($i\ne j$)\footnote{%
For the asqtad action, the $O_{12}$ part of $X_{ij}$ 
in eq.~(\ref{eq:mixingcoeffs}) corrects
an error in eqs.~(20-23) of Ref.~\cite{wlee-02}. 
The numerical impact of this error is, however, minor:
results for the mixing coefficients given
in Table II of Ref.~\cite{wlee-02} are changed by
less than $10^{-3}$.
We stress that for the HYP action,
the expressions and numerical values 
given in Ref.~\cite{wlee-02} are correct.
}
%
%
\begin{equation}
  \begin{split}
  X_{ij} = - \int_k 2 B F^2 s_1 s_2 
  &\big(s_1^N s_2^N P_3 \bar{c}_3^2 V^{\rm mix}_{P;ij} \\
  &+ s_1 s_3^N O_{12} V^{\rm mix}_{O;ij}
  \big) \,.
  \end{split}
  \label{eq:mixingcoeffs}
\end{equation}
The vertex factors for the cases of non-vanishing
mixing are collected in Table~\ref{tab:mixingcoeffs}.

\begin{table*}[htbp!]
\caption{ Vertex factors for non-vanishing bilinear mixing coefficients,
 needed in eq.~(\protect\ref{eq:mixingcoeffs}).
The components
$\mu$, $\nu$ and $\rho$ are all different, but otherwise arbitrary.
We use the shorthand $c_\mu=\cos(k_\mu)$.
Mixing coefficients for which both $i$ and $j$ are multiplied by
$(\gamma_5\otimes\xi_5)$ are the same, and are not shown separately.
  \label{tab:mixingcoeffs}
}
\begin{ruledtabular}
\begin{tabular}{cccc}
Operator-$i$ & Operator-$j$ &
$V^{\rm mix}_{P;ij}$ & $V^{\rm mix}_{O;ij}$
\\
\hline
$(\gamma_\mu\otimes\xi_\nu)$ & $(\gamma_\mu\otimes\xi_\mu)$  
& $2$ & $2[s_2 s_3^N - 2 s_2^N s_3]$ 
\\
$(\gamma_\mu\otimes\xi_{\mu5})$ & $(\gamma_\mu\otimes\xi_{\nu5})$ 
& $-2c_3c_4$ & $-2c_4[s_2 s_3^N c_3 - 2 s_2^N s_3 c_2]$ 
\\
$(\gamma_\mu\otimes\xi_{\mu\nu5})$ & $(\gamma_\mu\otimes\xi_{\rho\nu5})$ 
& $-[c_3+c_4]$ & $-s_2 s_3^N [c_3 + c_4] + 2 s_2^N s_3 [c_2 + c_4]$ 
\\
$(\gamma_{\mu\nu}\otimes\xi_{\mu5})$ & $(\gamma_{\mu\nu}\otimes\xi_{\rho5})$
& $c_4-c_3$ & $-s_2 s_3^N [c_3 - c_4] + 2 s_2^N s_3 [c_2 - c_4]$ \\
\end{tabular}
\end{ruledtabular}
\end{table*}

The contribution of $Y$-diagrams depends only on the
distance $\Delta=\sum_{\mu}(S-F)_\mu^2$. It vanishes
for $\Delta=0$, and is otherwise
\begin{equation}
Y_\Delta = \sum_{k=1}^{\Delta} I_k \,,\qquad
(\Delta\ge1)\,,
\end{equation}
where
\begin{equation}
I_\Delta = \int_k BF\left(s_1 s_1^N P_1 
+12 \bar{s}_1^2 s_2 s_2^N O_{21}\right) V_Y(\Delta)
\end{equation}
with vertex factors
\begin{equation}
  \begin{split}
  V_Y(1) & = 1 \,, \\
  V_Y(2) & = \frac{c_2 + c_3 + c_4}{3} \,, \\
  V_Y(3) & = \frac{c_2 c_3 + c_2 c_4 + c_3 c_4}{3} \,, \\
  V_Y(4) & = c_2 c_3 c_4 \,.
  \end{split}
\end{equation}

The tadpole contribution also depends only on the distance
$\Delta$. It is conveniently
divided into the contribution from gluon propagators
beginning and ending on the same smeared link,
$T_\Delta^a$, and the remainder, $T_\Delta^b$, which requires
$\Delta\ge2$.
The former is naturally combined with the self-energy
tadpole to yield
\begin{equation}
T_\Delta^a +ZT = (\Delta-1) \left[
I_{\rm MF} - \int_k (B/2) P_1\right]
\,.
\label{eq:type-a-tadpole}
\end{equation}
Here $I_{\rm MF}$ is present if mean-field improvement
is implemented, and is given by
\begin{eqnarray}
%
%
I_{\rm MF} = \int_k B \bar{s}_2^2 \left[
P_1 - 4 \bar{s}_1^2 O_{12}\right]
\,.
\label{eq:IMF-HYP}
\end{eqnarray}
Note that for unimproved staggered fermions, mean-field
improvement is commonly called tadpole improvement.
The numerical values of $I_{\rm MF}$ are
$\pi^2=9.869605$, $7.229736$ for unimproved staggered fermions with Wilson and
improved gauge actions, respectively, and
$1.053786$, $0.722795$ for HYP fermions with the same two gauge actions.

The second part of the tadpole contribution is unaffected
by mean-field improvement, and is
\begin{equation}
T_\Delta^b = \int_k 4B \bar{s}_1^2 \bar{s}_2^2 O_{12} V_T(\Delta)
\,,
\label{eq:type-b-tadpole}
\end{equation}
with vertex factors $V_T(0)=V_T(1)=0$ and
\begin{equation}
V_T(2)=1\,,\
V_T(3)=2+c_3\,,\
V_T(4)=3+2c_3+c_3 c_4\,.
\end{equation}

Finally, the non-tadpole self-energy contribution can be
obtained from the conservation of the taste-singlet vector
current:
\begin{equation}
Z = - X_{ii} - Y_1\,,\qquad
i=(\gamma_\mu\otimes 1)\,.
\end{equation}

These results hold for the HYP action with 
different choices for smearing kernel (entering through the
coefficients $d_{1-4}$)
and different choices of gauge action
(entering through the coefficients $c_{0-3}$ in the gluon
propagator). 

\section{One-loop results for asqtad fermions
 \label{app:one-loop-asqtad}}

Using asqtad rather than HYP-smeared fermions leads to
three changes: (i) the links are now $O(a^2)$ improved,
rather than HYP-smeared; (ii) the Naik term is present;
and (iii) the hypercube vector current is no longer conserved.
The impact of these changes is that, while the
X- and Y-diagrams can be obtained by simple
substitutions from those for HYP-smeared fermions, 
the tadpole and self-energy contributions 
must be calculated anew.
In detail, the changes from the previous section
are as follows:
\begin{itemize}
\item
The coefficients in the smearing kernel are now
$d_1=0$, $d_2=d_3=d_4=1$. These enter through $D_\mu$
and $\tilde{G}_{\nu,\mu}$.
\item
In all expressions,
$s_\mu^N$ now differs from $s_\mu$ due to the effect of the
Naik term on the propagator:
\begin{equation}
s_\mu^N = s_\mu (1+s_\mu^2/6)\,.
\end{equation}
\item
The form of the result for X-diagrams remains unchanged,
but $P_\mu$ and $O_{\mu\nu}$ are changed because of
the impact of the Naik term on the quark-gluon vertex.
They are replaced by $P^{NN}_\mu$ and $O^{NN}_{\mu\nu}$,
obtained from
\begin{eqnarray}
\CT^{NN}_{\mu\nu} &=& 
\sum_{\alpha,\beta} h^N_{\mu\alpha} h^N_{\nu\beta}
\CD^{\rm Imp}_{\alpha\beta} 
\\
&=&
B \left[\delta_{\mu\nu} P^{NN}_\mu +
(1-\delta_{\mu\nu})4{\bar s}_\mu{\bar s}_\nu O^{NN}_{\mu\nu}\right]
\,,
\label{eq:POdefNN}
\end{eqnarray}
where
\begin{equation}
h_{\mu\nu}^N = h_{\mu\nu} + \delta_{\mu\nu} s_\mu^2/6\,.
\end{equation}
\item
The form of the result from Y-diagrams is also unchanged, 
but now $P_\mu$ and $O_{\mu\nu}$ must be replaced by
$P^N_\mu$ and $O^N_{\mu\nu}$, which are obtained from
\begin{eqnarray}
\CT^{N}_{\mu\nu} &=& 
\sum_{\alpha,\beta} h^N_{\mu\alpha} h_{\nu\beta}
\CD^{\rm Imp}_{\alpha\beta} 
\label{eq:CTN}
\\
&=&
B \left[\delta_{\mu\nu} P^N_\mu +
(1-\delta_{\mu\nu})4{\bar s}_\mu{\bar s}_\nu O^N_{\mu\nu}\right]
\,.
\label{eq:POdefN}
\end{eqnarray}
These asymmetrical changes reflect the fact that the Naik term
enters when the gluon attaches to the external fermion leg but
not when it attaches to the operator. 
\item
Type-(b) tadpole diagrams are unchanged in form and
involve $O_{\mu\nu}$ without any change from the Naik term.
In other words, one uses eq.~(\ref{eq:type-b-tadpole}) with
$O_{\mu\nu}$ from eq.~(\ref{eq:POdef}).
\item
Type-(a) tadpole diagrams are replaced by
\begin{eqnarray}
T^a_\Delta +Z_T &=& (\Delta-1)\left[
\frac14
\int_k {\cal D}^{\rm Imp}_{11} (-5 + 3 c_1 + 3 c_2 - 3 c_1 c_2)\right.
\nonumber\\
&& \left.\qquad\qquad
+ 12 \int_k \tilde{\cal D}^{\rm Imp}_{12} (\bar s_1)^4 (\bar s_2)^2
+ \frac52 T^{Sym}  
\right]
\nonumber\\
&& +\frac14 \left[T^{Sym} - \int {\cal D}^{\rm Imp}_{11} c_1 (1+c_1)\right]
\,,
\label{eq:Ta}
\end{eqnarray}
where the $T^{Sym}$ terms arise from the tadpole-improvement 
of the links, with
\begin{equation}
T^{Sym} = \int_k \left({\cal D}^{\rm Imp}_{11} (\bar s_2)^2 - 
4 \tilde{\cal D}^{\rm Imp}_{12} (\bar s_1)^2(\bar s_2)^2 \right)
\,.
\label{eq:TSym}
\end{equation}
For the Wilson gauge action $T^{Sym}=\pi^2$, but this factor
is reduced for the improved gauge action to $7.229736$.
\item
The non-tadpole self-energy is given by
\begin{equation}
Z = \int_k (B^2 + B F I_z) \,,
\end{equation}
where 
\begin{eqnarray}
I_Z &=& c_1^N\left[1-2 (s_1^N)^2 F\right]
\left[\bar c_1^2 P_1^{NN}-3\bar c_2^2 P_2^{NN}\right]
\nonumber\\
&& -s_1s^N_1 P_1^{NN}  -12\bar s_1^2 s_2 s^N_2 O_{12}^{NN}
\nonumber\\
&& -s_1s^N_1(s_1^2/3-\bar c_1^2)D'_1
\nonumber\\
&&+ s_1^2 s_2 s_2^N (3/4-\bar{s}_1^2) \tilde G'_{1,2}
\nonumber\\
&& -12s_1 s^N_1 c_1^Ns_2s^N_2 F O^{NN}_{12}
\,.
\end{eqnarray}
Here $c_\mu^N=c_\mu(1+s_\mu^2/2)$, while
the new quantities $D'_\mu$ and $\tilde{G}'_{\nu,\mu}$
arise from the propagator from a smeared link to a thin (Naik) link.
They are defined by
\begin{equation}
\sum_\rho h^N_{\mu\rho} \CD^{\rm Imp}_{\rho\nu}
\equiv B \left[\delta_{\mu\nu} D'_\mu
+ (1-\delta_{\mu\nu}) \bar{s}_\mu \bar{s}_\nu \tilde{G}'_{\nu,\mu}
\right]\,.
\end{equation}
\end{itemize}

As noted in the main text, we have also calculated
the matching factor for the asqtad conserved vector current.
This is given by adding 1- and 3-link terms
to the hypercube current, both containing tadpole-improved
thin links:
\begin{eqnarray}
\lefteqn{V_\mu^{\rm CVC}(y) = 
[V_\mu\times S](y; W) +\frac1{8u_0} [V_\mu\times S](y; U)}
\nonumber\\
&&
 - \frac1{24}\times \frac1{16u_0^3}\sum_{\vec A} \eta_\mu(y)\times
\nonumber\\
&&
\bigg[
\bar\chi_{y\!+\!\vec A-2\hat\mu} 
U_\mu(y\!+\!\vec A\!-\!2\hat\mu,y\!+\!\vec A\!+\!\hat\mu)
\chi_{y\!+\!\vec A\!+\!\hat\mu}
\nonumber\\
&&+
\bar\chi_{y\!+\!\vec A\!-\!\hat\mu} 
U_\mu(y\!+\!\vec A\!-\!\hat\mu,y\!+\!\vec A\!+\!2\hat\mu)
\chi_{y\!+\!\vec A\!+\!2\hat\mu}
\nonumber\\
&&+
\bar\chi_{y\!+\!\vec A} 
U_\mu(y\!+\!\vec A,y\!+\!\vec A\!+\!3\hat\mu)
\chi_{y\!+\!\vec A\!+\!3\hat\mu}
+ h.c.)\bigg]\,,
\label{eq:CVC}
\end{eqnarray}
where the notation for bilinears is as in
eq.~(\ref{eq:bilindef}), except that the
second argument of $[V_\mu\times S]$ indicates the
type of links used to create the parallel transporter.
In addition, $\vec A$ is a vector running over the 8 positions
of the cube perpendicular to $\mu$, while
``h.c.'' implies interchange of the positions of $\bar\chi$
and $\chi$ fields and hermitian conjugation of the gauge fields.

At tree-level, and for physical external momenta,
the extra 1- and 3-link terms in the current cancel.
At one-loop, however, these terms lead to additional contributions.
For X-diagrams, the effect is to change the
vertex functions as follows:
\begin{equation}
V_i(k) \longrightarrow V_i(k) + \frac18\left[V_i(k)-V_i(3k)\right]\,.
\end{equation}
For Y-diagrams, the expression for ${\cal T}^N_{\mu\nu}$
in (\ref{eq:CTN}) is changed by the substitution
\begin{equation}
h_{\nu\beta} \longrightarrow h_{\nu\beta} - \delta_{\nu\beta}
c_\nu(1+c_\nu)/2\,,
\end{equation}
(with $h^N_{\mu\alpha}$ unchanged).
The T diagrams must be calculated anew, but turn out to exactly
cancel the ZT contribution (for any choice of gluon propagator).

The net result, as we have checked analytically, is that
the contributions of the X, Y and Z diagrams cancel exactly.
This provides an important check on our result for the Z diagram
with asqtad fermions.

\bibliographystyle{apsrev} 
\bibliography{ref} 

\end{document}